\documentclass[twocolumn,showpacs,prl]{revtex4}

\usepackage{graphicx}
\usepackage{dcolumn}
\usepackage{bm}

\begin{document}

\title{First order transition from correlated electron semiconductor 
to ferromagnetic metal in single crystalline FeSi$_{1-x}$Ge$_{x}$}

\author{ S. Yeo$^{1}$, S. Nakatsuji$^{1}$, A.D. Bianchi$^{2}$, 
P. Schlottmann$^{3}$, Z. Fisk$^{1,3}$, L. Balicas$^{1}$, P.A. 
Stampe$^{4}$, and R.J. Kennedy$^{4}$ }
\affiliation{
$^1$ National High Magnetic Field Laboratory (NHMFL), Tallahassee, Florida 32310\\
$^2$ Los Alamos National Laboratory, Los Alamos, New Mexico 87545\\
$^3$ Department of Physics, Florida State University, Tallahassee,
Florida 32306\\
$^4$ Department of Physics, Florida A\&M University, Tallahassee,
Florida 32307}

\date{\today}

\begin{abstract}
The phase diagram of FeSi$_{1-x}$Ge$_{x}$, obtained from magnetic, 
thermal and transport measurements on single crystals, shows a 
first-order transition from a correlated electron semiconductor to a ferromagnetic 
metal at a critical concentration, $x_{\rm c} \approx 0.25$. The gap of 
the insulating phase strongly decreases with $x$. The specific 
heat $\gamma$ coefficient appears to track 
the density of states of a Kondo insulator. The phase diagram is 
consistent with a correlation induced insulator-metal transition
in conjunction with disorder on the Si/Ge ligand site. 

\end{abstract}
\pacs{PACS numbers: 71.30.+h, 72.15.Rn, 75.50.Pp}
\maketitle

Kondo insulators (KIs) are Ce, Yb or U based small-gap semiconductors 
\cite{AF}, e.g. CeNiSn, Ce$_3$Bi$_4$Pt$_3$, and YbB$_{12}$. Many KIs 
are nonmagnetic with a Van-Vleck-like low-$T$ susceptibility, but 
an exponentially activated low-$T$ resistivity and electronic 
specific heat. Most KIs are not perfect semiconductors, because the 
hybridization gap is frequently only a pseudogap and/or there are 
intrinsic or impurity states in the band gap \cite{Rise}. A magnetic 
field gradually closes the gap yielding a metallic state for fields 
larger than a critical one \cite{Millis,Jaime}. 

FeSi is known as a correlated electron semiconductor \cite {jac} 
with an activation gap of about 0.05 eV and exhibits properties 
similar to those of KIs. Neutron scattering and M\"ossbauer 
experiments reveal no long-range magnetic order in FeSi \cite{wat,wer}. 
FeGe, on the other hand, is a long-range spiral metallic ferromagnet 
(with period of 700 $\mathring{\rm A}$ below 280K) with a saturated 
magnetic moment of about 1 $\mu_B$ in a magnetic field larger than 
0.3 T \cite{Lebech}. By alloying these two systems it is then 
possible to study the {\it transition from correlated electron 
semiconductor to ferromagnetic metal} (FM), with the following 
unique advantages: (i) Both compounds form in the same slightly 
distorted rocksalt structure and are soluble over the entire 
concentration range. (ii) Single crystals can be grown for the two 
end-compounds {\it and} the entire alloy range (previous studies 
were on polycrystalline samples and over a limited range of $x$ 
\cite{Reichl, Mani}). (iii) The substitution is on the sites of the 
ligand atoms, yielding a reduction of the gap, and not of the 
magnetically active ion (Fe), which would give rise to an impurity 
band in the gap \cite{schl}. (iv) The Kondo gap is large (about 50 
times larger than for other KIs) and therefore the system is less 
sensitive to internal strains and other impurities. (v) Since the 
spin-orbit coupling of Fe is much smaller than for rare earth and 
actinide ions, the orbital momentum is quenched and the magnetic 
susceptibility is exponentially activated, rather than Van-Vleck-like, 
and a useful tool to study the gap.  

Optical conductivity measurements indicate that FeSi is a strongly 
correlated system \cite{schles}, because (1) the gap disappears at 
unusually low $T$ (above 200 K) relative to its magnitude, 
and (2) the conductivity displaced from the gap region at low $T$ 
does not appear just above the gap, but is spread over a wide
energy range. The results are consistent with the usual KI 
picture. However, the gap is much larger than usual Kondo 
temperatures, so that FeSi could also be considered a mixed-valence 
insulator \cite{Varma}. Based on the spin-fluctuation theory, FeSi
has been argued a nearly ferromagnetic semiconductor \cite{tak}.
LDA$+U$ band calculations \cite{Anisimov} also suggest that FeSi is 
close to a magnetic instability. Increasing Hubbard $U$ leads to a first 
order transition from a strongly correlated semiconductor to a FM. 

Similarly, as a function of $U/V$, the half-filled Anderson lattice 
undergoes a first order transition from KI to a FM ordered state
\cite{DS}. Here $V$ is the hybridization of the Fe 3d states with 
the ligand atoms. According to X-ray photoelectron spectroscopy 
(XPS) studies and band calculations \cite{XPS}, $V$ is smaller for 
FeGe than for FeSi. Alloying these two systems may then lead to a 
first order transition from KI to FM. Here we present magnetic, 
transport, and thermal properties of FeSi$_{1-x}$Ge$_{x}$ to study 
the evolution of this transition.

Single crystals were grown by vapor transport method described in
Ref.~\cite{Richard}. Polycrystalline samples prepared by arcmelting
of the stoichiometric mixture of elements were sealed in an evacuated
quartz tube with the chemical agent iodine and heated in a home made
two zone furnace for a week. The temperature variations on both sides 
of the tube were measured with a commercial thermometer and found 
to be less than 5K. The Ge/Si concentration was determined by 
Energy Dispersive X-ray analysis (the uncertainty is $\pm 5\%$).

All measurements were performed using the same set of crystals. The 
DC susceptibility was obtained with a Quantum Design MPMS between 
2 K and 350 K, the resistivity with a standard four-probe 
technique in the temperature range from 2 K to 300 K, and the 
specific heat $C_p(T)$ with a relaxation method down to 0.35 K.

\begin{figure}[tb]
\includegraphics[width=8.5cm]{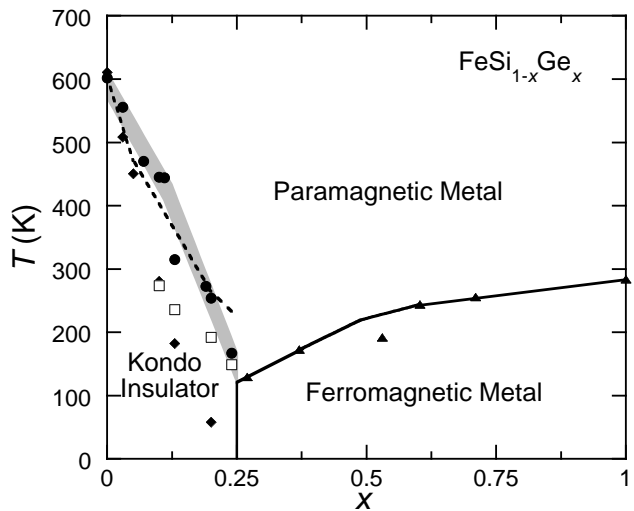}
\caption{Phase diagram of FeSi$_{1-x}$Ge$_{x}$. Solid circles 
and solid diamonds represent gaps from $\chi(T)$ and $\rho(T)$, 
respectively, open squares correspond to the minimum temperature
of $\rho(T)$, and the solid triangles are the ferromagnetic 
$T_{\rm C}$. The solid lines and shaded area are guides to the eyes 
for phase boundaries and energy gap, respectively. 
The dashed line is the decrease of the gap with $x$ according to a 
selfconsistent theory for KI with ligand defects.}
\label{phase}
\end{figure}

Fig.~\ref{phase} displays the phase diagram of FeSi$_{1-x}$Ge$_{x}$
obtained from magnetic, transport and thermal measurements. The 
solid circles, solid diamonds, and open squares represent the gap 
in the susceptibility, the transport gap and the resistivity 
minimum for $x \leq 0.24$, respectively. The triangles correspond 
to the ferromagnetic transition temperature $T_{\rm C}$ for $x > 
0.25$. Note the first order transition from nonmagnetic semiconductor 
to FM at $x \approx 0.25$.

\begin{figure}[tb]
\includegraphics[width=8.5cm]{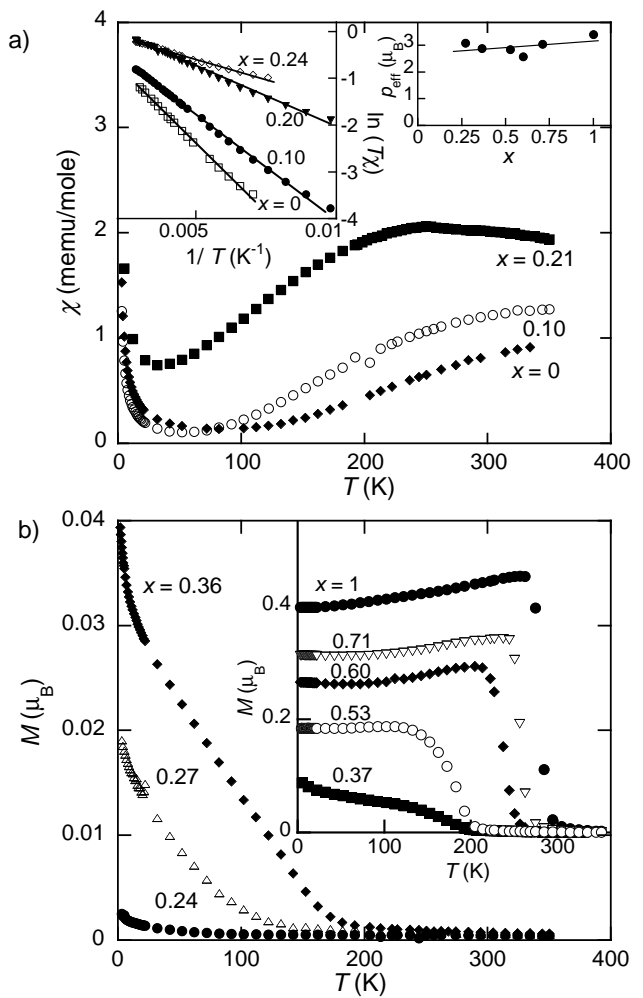}
\caption{(a) Susceptibility vs. temperature for $x=0, 0.1$, and 0.21 
measured in a field of 0.1 T. The slope in the left inset is the 
thermal activation energy for $\chi(T)$ for $x \leq$ 0.24. The 
right inset represents the effective moment for $x > 0.25$. The 
solid lines are guides to the eye. (b) Magnetization as a function 
of $T$ in a field of 0.1 T for $x \ge 0.25$. For $x > 0.6$ the 
transition is discontinuous, while for $x < 0.6$ the ferromagnetism 
disappears continuously.}
\label{mag}
\end{figure}

Fig.~\ref{mag} (a) and (b) show the temperature dependence of the
magnetic susceptibility of FeSi$_{1-x}$Ge$_{x}$ for $x \leq 0.24$ 
and $x \geq 0.24$, respectively. For $x \leq 0.24$ $\chi(T)$ 
increases upon heating following a thermally activated Curie law,
\begin{equation}
\chi(T) = (C/T) \exp(- \Delta_s / k_B T ) \ ,
\label{chiactivated}
\end {equation}
where $\Delta_{s}$ is the spin gap. The left inset of Fig.~\ref{mag} 
(a) shows that $\ln(T \chi(T)$) vs. $T^{-1}$ for $150 K < T < 350 K$
follows a straight line, indicating that Eq.~(\ref{chiactivated}) 
represents $\chi(T)$ well. The most significant effects of the Ge 
substitution on $\chi(T)$ are: (i) a systematic increase with $x$, 
(ii) a maximum starts to show at about 300 K for $x \geq 0.2$, 
and (iii) the spin gap $\Delta_s$ monotonically decreases with $x$, 
as seen in Fig.~\ref{phase}. At low $T$, $\chi(T)$ displays a Curie 
tail, which is attributed to impurities or defects. Since this tail 
is approximately the same for all $x \le 0.24$, we conclude that 
the impurity concentration is roughly constant with $x$.

In contrast, $\chi(T)$ qualitatively changes with further Ge doping. 
For $0.25 \le x \le 0.50$ the magnetization gradually increases 
with decreasing $T$ as seen in Fig.~\ref{mag} (b), suggesting the 
appearance of a ferromagnetic component. In fact, for $x \geq 0.53$, 
the ferromagnetic transition is given by a step-like increase upon 
cooling (inset of Fig.~\ref{mag} (b)). $T_{\rm C}$ is determined 
by the kink in the magnetization and shown as the solid triangles in 
Fig.~\ref{phase}. $T_{\rm C}$ monotonically increases with $x$ until 
it reaches the known value for FeGe \cite{Lebech}. The sharp jump of 
the magnetization of FeGe around 280 K appears to be consistent 
with a first order transition \cite{bak}. The distorted rocksalt 
structure lacks of inversion symmetry giving rise to a 
Dzyaloshinskii-Moriya (DM) interaction. A renormalization group study 
\cite{bak} of the ferromagnetic DM instability
predicts that the magnetic structure is helical of long period and 
the transition into the paramagnet is of first order. Experimentally
the transition remains discontinuous for $x > 0.6$. In the range 
$0.25 < x < 0.60$, on the other hand, the magnetization tends to 
zero continuously at $T_{\rm C}$, probably as a consequence of 
disorder in the sample. The alloying of FeSi with FeGe necessarily 
leads to Ge-rich and Si-rich regions and hence to a distribution of 
transition temperatures, such that the total magnetization is 
continuous. The effective moment ($p_{eff}$)¥ of Fe is obtained by fitting a 
Curie-Weiss law for $T \ge T_{\rm C}$, and is shown in the right 
inset of Fig.~\ref{mag} (a). $p_{eff}$¥ are almost independent of $x$ and roughly correspond to 
$S = 1$. 

\begin{figure}[tb]
\includegraphics[width=8.5cm]{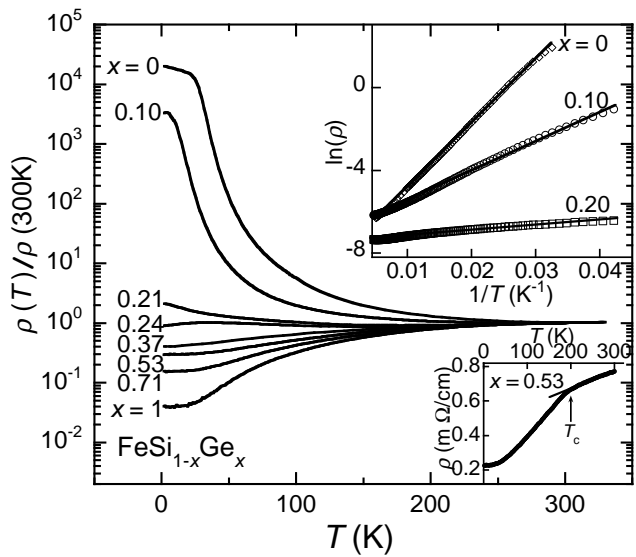}
\caption{$\rho(T)/\rho(300 {\rm K})$ for FeSi$_{1-x}$Ge$_{x}$. 
The slope in the inset corresponds to the thermal activation gap 
for $x \leq 0.21$.}
\label{res}
\end{figure}

The $T$-dependence of the resistivity, $\rho(T)$, normalized to 
its value at 300 K is shown in Fig.~\ref{res}. Two distinct 
regimes have to be distinguished: For $x < 0.24$ the resistivity 
decreases with $T$, characteristic of a semiconductor, while for 
$x > 0.25$ $\rho(T)$ is an increasing function and the behavior 
is metallic. Starting from the nonmagnetic semiconductor FeSi, 
the Ge substitution dramatically reduces $\rho(T)$. We assume 
a thermal activation law describes the insulating behavior in 
the Si rich region,
\begin{equation}
\rho(T) = \rho_{0} \exp(\Delta_t /2 k_B T) \ ,
\label{rhoactivated}
\end {equation}
where $\Delta_t$ is the transport gap. For $x \le 0.21$ the 
resistivity for $25$ K $< T < 210$ K is well represented by the 
activation law (see inset of Fig.~\ref{res}). The dependence 
of $\Delta_t$ on $x$ is displayed in Fig.~\ref{phase}. For FeSi 
$\Delta_s$ and $\Delta_t$ have the same value, but with increasing 
$x$ the transport gap becomes much smaller than the magnetic spin 
gap. This difference is due to the randomness in the sample, 
consequence of the alloying. While in a transport measurement the 
electrons travel along the path of least resistance, i.e. the one 
with the smallest effective gap, a thermodynamic measurement 
averages over the entire sample. Necessarily, $\rho(T)$ should 
have the smaller gap. Below 20 K the resistivity saturates, which 
is attributed to impurity states in the gap.

The resistivity of FeGe is clearly metallic. Substituting Si for Ge
gradually increases $\rho(T=0)$ because of disorder scattering, but
the system remains a poor metal up to the metal-insulator transition 
at $x_c \approx 0.25$. For $x \geq 0.53$ a small kink in $\rho(T)$ 
can be observed at the onset of ferromagnetic long-range order at 
$T_{\rm C}$. The qualitative change in $\chi(T)$ and $\rho(T)$ at 
$x_{\rm c} = 0.25$ is consistent with a first order transition. 
For $x < x_{\rm c}$ they follow activation laws with their gaps 
decreasing with $x$, while for $x > x_{\rm c}$ the system is FM 
at low $T$.

\begin{figure}[tb]
\includegraphics[width=8.5cm]{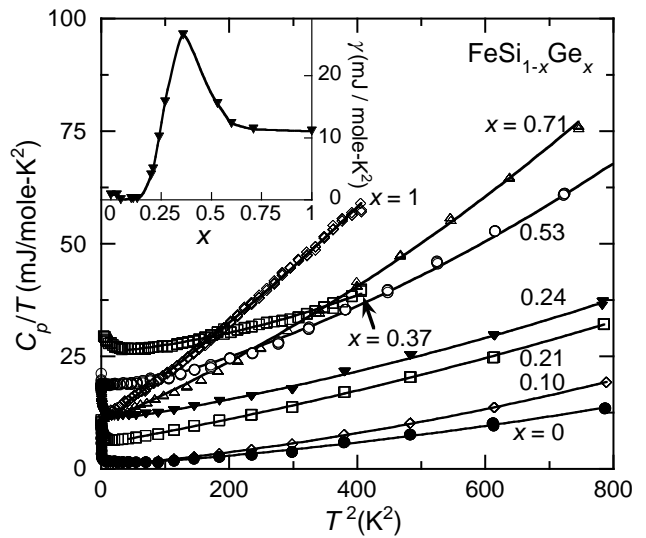}
\caption{Heat capacity data for FeSi$_{1-x}$Ge$_{x}$ shown as 
$C_{p}(T)/T$ vs. $T^{2}$. The inset displays the $\gamma$ values.}
\label{gamma}
\end{figure}

The specific heat divided by temperature, $C_p(T)/T$, vs. $T^2$ 
is shown in Fig.~\ref{gamma}. For $T < 2 K$, $C_p/T$ displays an 
upturn, believed to originate from 1-2\% of impurities. The 
estimated low-$T$ entropy up to 2~K is consistent with the 
Curie constant obtained from the tail of the impurity states of 
$\chi(T)$ for $x \le 0.24$. The specific heat of the compound 
(electrons and phonons) is obtained by fitting the data for $5~K
< T < 40~K$ to the expression
\begin{equation}
C_p(T) = \gamma T + \beta T^3 + \delta T^5 \ ,
\label{Cp}
\end{equation}
where $\gamma T$ is the electronic contribution, shown in
the inset of Fig.~\ref{gamma}. In the insulating regime, i.e. $x < 
x_{\rm c}$, $\gamma$ is almost zero, suggesting a very small or no 
density of states at the Fermi energy. However, for $x > x_{\rm c}$, 
i.e. in the FM regime, $\gamma$ is nonzero, has its largest value at $x 
\approx 0.4$ after a steep increase, and then it decreases to a 
constant value (for $x \geq$ 0.6) of about 11 mJ/mole-K$^2$. The
Debye temperature can be obtained from $\beta$. $\theta_D$ decreases
with $x$ from 600~K for FeSi to 280~K for FeGe.

As seen in Fig.~\ref{phase}, FeSi$_{1-x}$Ge$_{x}$ has a first order 
transition from KI to FM at $x_c \approx 0.25$. Recent LDA$+U$ 
band structure calculations \cite{Anisimov} have shown that FeSi 
is a semiconductor that is close to a ferromagnetic instability. With 
increasing $U$, it is found that at a critical $U_c = 3.2$ eV, a first 
order transition from paramagnetic semiconductor to FM takes place. 
The calculation \cite{Anisimov1} has been extended to the alloy 
FeSi$_{1-x}$Ge$_{x}$ using experimental lattice constants and 
averages of Si and Ge for the potential parameters. A first order 
insulator-metal transition is predicted at $x_c = 0.4$ for $U =3.7$ 
eV, in agreement with the experimental findings.

Detailed LDA band calculations by Mattheiss and Hamann \cite{mat} 
show that FeSi is a small (indirect) gap semiconductor.
The bands closest to the Fermi level have 
predominantly Fe 3d character with some portions of weak dispersion 
about the $\Gamma$ and $X$ points. All other bands are either filled 
or empty for the temperature range under consideration. It should be then 
meaningful to discuss FeSi$_{1-x}$Ge$_{x}$ within the framework 
of the Anderson lattice Hamiltonian without orbital degeneracy (one 
band). The magnetic instabilities of a KI (two electrons per site) 
have been studied within a mean-field slave-boson approach \cite{DS}.
The key parameter driving 
the magnetic phases is the Hubbard/Anderson $U$ over the 
hybridization between the localized and conduction states $V$. 
(In the LDA$+U$ approach \cite{Anisimov,Anisimov1} the correlation 
strength is $U$ over the bandwidth $W$). Two magnetic phases, an 
antiferromagnetic and a ferromagnetic one, are predicted. With 
increasing $U/V$, the paramagnetic KI first undergoes a second order 
transition into an antiferromagnetic insulating phase. This, 
however, requires a bipartite lattice (two interpenetrating 
sublattices) symmetry, which is absent for the rocksalt structure. 
Consequently, antiferromagnetism is suppressed and the system has 
a discontinuous transition to the FM phase \cite{DS}.

For FeSi$_{1-x}$Ge$_{x}$, we consider $U$ a property of the Fe 3d 
shell and a constant for the alloy series. The hybridization of 
the 3d states with the ligand atoms, however, depends on $x$ and 
the simplest assumption is an average hybridization, i.e. $V_{eff} 
= (1-x) V_{Si} + x V_{Ge}$. According to XPS and band calculations 
for FeSi and FeGe \cite{XPS}, $V_{Si}$ is larger than $V_{Ge}$. 
Hence, the effective hybridization, $V_{eff}$, decreases and 
$U/V_{eff}$ increases with $x$, thus leading to a discontinuous 
transition at $(U/V)_{cr}$. The parameters $U$ and $V_{eff}$ are 
such that the compound is actually in the mixed valence 
and not quite in the Kondo regime \cite{Varma}.Within the 
mean-field approach the critical parameter is $(U/V)_{cr} = 1.54$.

The $\gamma$ coefficient, shown in the inset of Fig.~\ref{gamma}, 
is a measure of the density of states at the Fermi level. Based 
on the KI picture, in the gapped region we have $\gamma = 0$, while 
in the metallic region $\gamma \neq 0$. The magnetization (and 
also $T_{\rm C}$) increases with $x$ (or $U/V_{eff}$), so that the 
number of electrons (holes) in the majority (minority) conduction 
(valence) bands increase with $x$. Hence, $\gamma$ as a function 
of $x$ probes the density of states of the KI, which has a peak 
above (below) the band edges and then reaches a constant value 
for larger energy ($x$), in agreement with Fig.~\ref{gamma}. A large 
magnetic field is expected to favor a magnetic state close to the 
insulator-metal transition\cite{Millis,Jaime,Anisimov1}.
However, we found that the required fields are larger than the 45 T 
available at the NHMFL.

Finally, we address the rate of decrease of the gap with $x$, which 
is by far too large to be explained by the $x$-dependence of $V_{eff}$. 
It is then necessary to invoke the disorder introduced by the Ge 
substitution. An impurity in a KI gives rise to a bound state in 
the gap. Two situations have to be distinguished: (i) Substitution 
of magnetically active ions (Fe, Kondo holes) yields an impurity 
band close to the center of the band, while (ii) if ligand atoms 
are replaced, tails of impurity states develop close to the band 
edges reducing the spin gap \cite{schl}. Assuming that Ge enters the 
lattice randomly on ligand sites a selfconsistent calculation 
\cite{schl} yields a much faster decrease of the gap with $x$ as shown 
in Fig.~\ref{phase}
(normalized to the gap for FeSi for $V_{Ge} = 0.67 V_{Si}$ and 
$U/V_{Si} = 1.5$).

In summary, we studied the magnetic, transport, and thermal properties 
of single crystalline FeSi$_{1-x}$Ge$_{x}$ and found a first order
transition from the correlated insulator phase to a ferromagnetic 
metal phase at 
$x\approx 0.25$. The systematic change of the spin and transport 
gaps in the insulating phase, and the evolution of the magnetization 
and the $\gamma$-coefficient of the specific heat in the metallic 
phase are consistent with the Kondo insulator picture.

This work was performed at the NHMFL, which 
is supported by NSF Cooperative Agreement No. DMR-9527035 and by 
the State of Florida. This work is also supported by NSF and DOE through grants 
Nos.DMR-9971348, DMR-0105431 and DE-FG02-98ER45797.

\end{document}